\documentclass[twocolumn]{aastex62}

\usepackage{graphics,graphicx,xspace,natbib,amssymb}
\usepackage[caption=false]{subfig}
\usepackage{amsmath}
\usepackage{threeparttable}
\usepackage{makecell}

\usepackage{multirow}
\usepackage{fontawesome5}
\usepackage[encapsulated]{CJK}

\newcommand{\comment}[1]{}

\let\itAA\AA
\renewcommand{\AA}{\mathrm{\itAA}}

\shorttitle{Medium Bands, Mega Science}
\shortauthors{Suess et al.}

\begin{document}

\title{Medium Bands, Mega Science: a JWST/NIRCam Medium-Band Imaging Survey of Abell 2744}
\shortauthors{Suess et al.}

\author[0000-0002-1714-1905]{Katherine A. Suess}
\altaffiliation{NHFP Hubble Fellow}
\affiliation{Kavli Institute for Particle Astrophysics and Cosmology and Department of Physics, Stanford University, Stanford, CA 94305, USA}


\author[0000-0003-1614-196X]{John R. Weaver}
\affiliation{Department of Astronomy, University of Massachusetts, Amherst, MA 01003, USA}

\author[0000-0002-0108-4176]{Sedona H. Price}
\affiliation{Department of Physics and Astronomy and PITT PACC, University of Pittsburgh, Pittsburgh, PA 15260, USA}

\author[0000-0002-9651-5716]{Richard Pan}
\affiliation{Department of Physics and Astronomy, Tufts University, 574 Boston Ave., Medford, MA 02155, USA}

\author[0000-0001-9269-5046]{Bingjie Wang (\begin{CJK*}{UTF8}{gbsn}王冰洁\ignorespacesafterend\end{CJK*})}
\affiliation{Department of Astronomy \& Astrophysics, The Pennsylvania State University, University Park, PA 16802, USA}
\affiliation{Institute for Computational \& Data Sciences, The Pennsylvania State University, University Park, PA 16802, USA}
\affiliation{Institute for Gravitation and the Cosmos, The Pennsylvania State University, University Park, PA 16802, USA}

\author[0000-0001-5063-8254]{Rachel Bezanson}
\affiliation{Department of Physics and Astronomy and PITT PACC, University of Pittsburgh, Pittsburgh, PA 15260, USA}

\author[0000-0003-2680-005X]{Gabriel Brammer}
\affiliation{Cosmic Dawn Center (DAWN), Copenhagen, Denmark}
\affiliation{Niels Bohr Institute, University of Copenhagen, Jagtvej 128, Copenhagen, Denmark}

\author[0000-0002-7031-2865]{Sam E. Cutler}
\affiliation{Department of Astronomy, University of Massachusetts, Amherst, MA 01003, USA}

\author[0000-0002-2057-5376]{Ivo Labb\'e}
\affiliation{Centre for Astrophysics and Supercomputing, Swinburne University of Technology, Melbourne, VIC 3122, Australia}

\author[0000-0001-6755-1315]{Joel Leja}
\affiliation{Department of Astronomy \& Astrophysics, The Pennsylvania State University, University Park, PA 16802, USA}
\affiliation{Institute for Computational \& Data Sciences, The Pennsylvania State University, University Park, PA 16802, USA}
\affiliation{Institute for Gravitation and the Cosmos, The Pennsylvania State University, University Park, PA 16802, USA}

\author[0000-0003-2919-7495]{Christina C.\ Williams}
\affiliation{NSF’s National Optical-Infrared Astronomy Research Laboratory, 950 North Cherry Avenue, Tucson, AZ 85719, USA}

\author[0000-0001-7160-3632]{Katherine E. Whitaker}
\affiliation{Department of Astronomy, University of Massachusetts, Amherst, MA 01003, USA}
\affiliation{Cosmic Dawn Center (DAWN), Denmark} 


\author[0000-0001-8460-1564]{Pratika Dayal}
\affiliation{Kapteyn Astronomical Institute, University of Groningen, 9700 AV Groningen, The Netherlands}

\author[0000-0002-2380-9801]{Anna de Graaff}
\affiliation{Max-Planck-Institut f\"ur Astronomie, K\"onigstuhl 17, D-69117, Heidelberg, Germany}

\author[0000-0002-1109-1919]{Robert Feldmann}
\affiliation{Department of Astrophysics, University of Zurich, Zurich, CH-8057, Switzerland}

\author[0000-0002-8871-3026]{Marijn Franx}
\affiliation{Leiden Observatory, Leiden University, P.O.Box 9513, NL-2300 AA Leiden, The Netherlands}

\author[0000-0001-7440-8832]{Yoshinobu Fudamoto} 
\affiliation{Center for Frontier Science, Chiba University, 1-33 Yayoi-cho, Inage-ku, Chiba 263-8522, Japan}

\author[0000-0001-7201-5066]{Seiji Fujimoto}\altaffiliation{NHFP Hubble Fellow}
\affiliation{Department of Astronomy, The University of Texas at Austin, Austin, TX 78712, USA}

\author[0000-0001-6278-032X]{Lukas J. Furtak}
\affiliation{Department of Physics, Ben-Gurion University of the Negev, P.O. Box 653, Be’er-Sheva 84105, Israel}

\author[0000-0003-4700-663X]{Andy D. Goulding}
\affiliation{Department of Astrophysical Sciences, Princeton University, 4 Ivy Lane, Princeton, NJ 08544, USA}

\author[0000-0002-5612-3427]{Jenny~E.~Greene}
\affiliation{Department of Astrophysical Sciences, Princeton University, 4 Ivy Lane, Princeton, NJ 08544, USA}

\author[0000-0002-3475-7648]{Gourav Khullar}
\affiliation{Department of Physics and Astronomy and PITT PACC, University of Pittsburgh, Pittsburgh, PA 15260, USA}

\author[0000-0002-5588-9156]{Vasily Kokorev}
\affiliation{Kapteyn Astronomical Institute, University of Groningen, 9700 AV Groningen, The Netherlands}

\author[0000-0002-7613-9872]{Mariska Kriek}
\affiliation{Leiden Observatory, Leiden University, P.O.Box 9513, NL-2300 AA Leiden, The Netherlands}

\author[0000-0002-5337-5856]{Brian Lorenz}
\affiliation{Department of Astronomy, University of California, Berkeley, CA 94720, USA}

\author[0000-0001-9002-3502]{Danilo Marchesini}
\affiliation{Department of Physics \& Astronomy, Tufts University, MA 02155, USA}

\author[0000-0003-0695-4414]{Michael V. Maseda}
\affiliation{Department of Astronomy, University of Wisconsin-Madison, 475 N. Charter St., Madison, WI 53706, USA}

\author[0000-0003-2871-127X]{Jorryt Matthee}
\affiliation{Institute of Science and Technology Austria (ISTA), Am Campus 1, 3400 Klosterneuburg, Austria}

\author[0000-0001-8367-6265]{Tim B. Miller}
\affiliation{Center for Interdisciplinary Exploration and Research in Astrophysics (CIERA), Northwestern University, 1800 Sherman Ave, Evanston IL 60201, USA}

\author[0000-0001-7300-9450]{Ikki Mitsuhashi}
\affiliation{Department of Astronomy, The University of Tokyo, 7-3-1 Hongo, Bunkyo, Tokyo 113-0033, Japan}

\author[0000-0002-8530-9765]{Lamiya A. Mowla}
\affiliation{Dunlap Institute for Astronomy \& Astrophysics, Toronto, Ontario, M5S 3H4, Canada}

\author[0000-0002-9330-9108]{Adam Muzzin}
\affiliation{Department of Physics and Astronomy, York University, 4700 Keele Street, Toronto, ON, M3J 1P3, Canada}

\author[0000-0003-2895-6218]{Rohan P.\ Naidu}
\altaffiliation{NHFP Hubble Fellow}
\affiliation{MIT Kavli Institute for Astrophysics and Space Research, 77 Massachusetts Ave., Cambridge, MA 02139, USA}

\author[0000-0003-2804-0648 ]{Themiya Nanayakkara}
\affiliation{Centre for Astrophysics and Supercomputing, Swinburne University of Technology, PO Box 218, Hawthorn, VIC 3122, Australia}

\author[0000-0002-7524-374X]{Erica J. Nelson}
\affiliation{Department for Astrophysical \& Planetary Science, University of Colorado, Boulder, CO 80309, USA}

\author[0000-0001-5851-6649]{Pascal A. Oesch}
\affiliation{Department of Astronomy, University of Geneva, Chemin Pegasi 51, 1290 Versoix, Switzerland}
\affiliation{Cosmic Dawn Center (DAWN), Copenhagen, Denmark}

\author[0000-0003-4075-7393]{David J. Setton}\thanks{Brinson Prize Fellow}
\affiliation{Department of Astrophysical Sciences, Princeton University, 4 Ivy Lane, Princeton, NJ 08544, USA}

\author[0009-0007-1787-2306]{Heath Shipley}
\affiliation{Department of Physics, Texas State University, San Marcos, TX 78666, USA}

\author[0000-0001-8034-7802]{Renske Smit}
\affiliation{ Astrophysics Research Institute, Liverpool John Moores University, 146 Brownlow Hill, Liverpool L3 5RF, UK}

\author[0000-0003-3256-5615]{Justin S. Spilker}
\affiliation{Department of Physics and Astronomy and George P. and Cynthia Woods Mitchell Institute for Fundamental Physics and Astronomy, Texas A\&M University, 4242 TAMU, College Station, TX 77843-4242}

\author[0000-0002-8282-9888]{Pieter van Dokkum}
\affiliation{Department of Astronomy, Yale University, 52 Hillhouse Ave., New Haven, CT, 06511, USA} 

\author[0000-0002-0350-4488]{Adi Zitrin}
\affiliation{Department of Physics, Ben-Gurion University of the Negev, P.O. Box 653, Be’er-Sheva 84105, Israel}


\email{suess@stanford.edu}

\begin{abstract}
In this paper, we describe the ``Medium Bands, Mega Science" JWST Cycle 2 survey (JWST-GO-4111) and demonstrate the power of these data to reveal both the spatially-integrated and spatially-resolved properties of galaxies from the local universe to the era of cosmic dawn. Executed in November 2023, MegaScience obtained $\sim30$arcmin$^2$ of deep multiband NIRCam imaging centered on the $z\sim0.3$ Abell 2744 cluster, including eleven medium-band filters and the two shortest-wavelength broad-band filters, F070W and F090W. Together, MegaScience and the UNCOVER Cycle 1 treasury program provide a complete set of deep ($\sim28-30 \textrm{mag}_{\rm{AB}}$) images in {\it all} NIRCam medium- and broad-band filters. This unique dataset allows us to precisely constrain photometric redshifts, map stellar populations and dust attenuation for large samples of distant galaxies, and examine the connection between galaxy structures and formation histories. MegaScience also includes $\sim17$ arcmin$^2$ of NIRISS parallel imaging in two broad-band and four medium-band filters from $0.9-4.8\mu$m, expanding the footprint where robust spectral energy distribution (SED) fitting is possible. We provide example SEDs and multi-band cutouts at a variety of redshifts, and use a catalog of JWST spectroscopic redshifts to show that MegaScience improves both the scatter and catastrophic outlier rate of photometric redshifts by factors of 2-3. Additionally, we demonstrate the spatially-resolved science enabled by MegaScience by presenting maps of the [OIII] line emission and continuum emission in three spectroscopically-confirmed $z>6$ galaxies. We show that line emission in reionization-era galaxies can be clumpy, extended, and spatially offset from continuum emission, implying that galaxy assembly histories are complex even at these early epochs.  
We publicly release fully reduced mosaics and photometric catalogs for both the NIRCam primary and NIRISS parallel fields (\href{https:/jwst-uncover.github.io/megascience/}{jwst-uncover.github.io/megascience}). 
\end{abstract}


\section{Introduction}

The {\it James Webb Space Telescope} allows us, for the first time, to map the rest-frame optical and infrared emission from faint galaxies at $z\gtrsim3$. In the first $\sim$1.5 years of JWST data, these new capabilities have already begun to transform our understanding of galaxy formation and evolution. Luminous $z>10$ galaxies appear to be $\gtrsim10$ times more common than expected from theoretical predictions \citep[e.g.,][]{naidu22lum,adams23,atek23,austin23,bradley23,casey23,finkelstein23,robertson23}. Galaxy formation may have begun more rapidly than expected, with massive ($\gtrsim10^{10}M_\odot$) galaxies reported very early in the universe ($z\gtrsim7$) \citep[e.g.,][]{labbe23_massive,boylan-kolchin23,akins23,xiao23} and massive quiescent galaxies emerging as early as $z\sim4.6$ \citep[e.g.,][]{carnall23,valentino23,setton24}. New classes of objects --- particularly ``little red dots", thought to be predominantly highly dust-obscured active galactic nuclei --- appear to be remarkably common in deep-field observations \citep[e.g.,][]{matthee23,labbe23_lrd,furtak23_triplylensed,greene23,williams23,barro23,kokorev23c,kokorev24,wang24_brd}. 

Due to the targeted nature of spectroscopy, many of these initial discoveries rely on accurate interpretation of broad-band photometry. Spectroscopy is also significantly more expensive than imaging, meaning that photometry is also often necessary to select statistically large samples. While spectral energy distribution (SED) modeling of broad-band photometry is able to efficiently infer the physical properties of large samples of galaxies, the coarse wavelength resolution can permit contradictory physical interpretations. For example, models of photometric data alone can confuse the Lyman and Balmer breaks, obtaining dramatically incorrect photometric redshift solutions \citep[e.g.,][]{donnan23,arrabal-haro23,naidu22,zavala23}. It can be nearly impossible to correctly disentangle the relative contribution of line and continuum emission to broad-band photometry, leading to incorrect photometric redshifts or biased stellar masses \citep[e.g.,][]{laporte23,sarrouh24}. Broad-band photometric modeling may fail to identify new classes of objects with physical properties that lie beyond the modeling assumptions. For example, standard SED fitting yielded high stellar masses and redshifts for the objects in \citet{labbe23_massive}, but later spectroscopy revealed many of these objects to be a new population of reddened AGN that were not included in our pre-JWST photometric redshift template sets \citep[e.g.,][]{kocevski23,greene23}. 

The task of accurately inferring the physical properties of galaxies from photometry is further complicated by ``outshining": young stars so completely dominate the integrated light from galaxies that it is easy to miss up to a factor of ten more mass hidden in low-luminosity old stars (e.g., \citealt{papovich01}; for the predicted impact on $z>7$ galaxies with JWST see \citealt{whitler23,narayanan24,wang24_sys}). Spatially-resolved SED fitting studies, which can begin to disentangle line- and continuum-dominated regions, often find up to 40\% larger stellar masses and 0.3~dex older ages than spatially-integrated studies \citep[e.g.,][]{zibetti09,wuyts12,sorba15}. These differences between integrated and resolved properties may be even more dramatic in the high-redshift universe, where high equivalent widths of 1000-3000$\AA$ are common and can bias stellar masses and ages by up to a full dex (e.g., \citealt{gimenez-arteaga23}; see also \citealt{perez-gonzalez23}).

Building an accurate picture of a large sample of galaxies in the distant universe with JWST therefore requires spatially-resolved observations that can distinguish between nebular line and stellar continuum emission. These spatially- and spectrally-resolved observations provide a viable pathway forward towards breaking these model degeneracies and thereby producing accurate redshifts, stellar masses, star formation rates, and other stellar population properties. The most efficient way to achieve these results is to exploit one of JWST/NIRCam's most innovative features: a suite of twelve medium bands covering the full 1.5-5$\mu$m range \citep{rieke23_nircam}. These medium-band filters are effective at mapping strong emission lines, but they also have the continuum sensitivity required to probe older stellar populations in line-free regions. Together, these medium-band data allow us to efficiently map both the mass and approximate formation time of thousands of galaxies simultaneously. 

Ground-based observations have long shown the power of optical and near-infrared medium bands to finely sample the Balmer break at $z\sim0-4$ and constrain the physical properties of distant galaxies (e.g., COMBO-17, \citealt{wolf03}; NMBS, \citealt{whitaker11}; SHARDS, \citealt{perez-gonzalez13}; ZFOURGE, \citealt{straatman16}; FENIKS, \citealt{esdaile21}). While most early JWST programs used only a single medium-band filter, F410M, several more recent programs have begun to leverage the power of NIRCam's multiple medium bands: the JWST Extragalactic Medium-band Survey includes five 1-4$\mu$m medium bands \citep{williams23b}; parallel observations conducted as part of the CAnadian NIRISS Unbiased Cluster Survey include nine 1-4$\mu$m medium bands, with the three missing medium bands forthcoming in PID 3362 (PI Muzzin), and the Jades Origin Field includes six 1-3$\mu$m medium bands \citep{eisenstein23_jof}. These datasets are already beginning to demonstrate the power of multiple medium bands to constrain the redshifts, stellar ages, masses, and emission line equivalent widths of very large samples of galaxies \citep[e.g.,][]{withers23}.  

Here we present the ``Medium Bands, Mega Science" survey, designed to image the full Abell 2744 cluster field with {\it all dozen} of JWST/NIRCam's medium bands. Combined with existing broad-band imaging from the Cycle 1 UNCOVER Treasury Program \citep{bezanson22}, these observations make up a complete set of deep images in all of NIRCam's broad- and medium-band filters. This 20-band JWST imaging results in spatially-resolved ``ultra low-resolution" $R\sim15$ spectrophotometry for all $\sim70,000$ sources in the field, with nearly continuous photometric coverage from $0.7-5\mu$m. 

Section~\ref{sec:observations} describes our survey design, data reduction, and cataloguing methods. Section~\ref{sec:science} describes our science goals, including both spatially-integrated science (including example SEDs; Section~\ref{sec:integrated-science}) and spatially-resolved science (including example multi-band color images; Section~\ref{sec:resolved_science}). Section~\ref{sec:demo} provides a demonstration of the power of our data to localize strong line emission in distant galaxies, presenting evidence for clumpy and offset [OIII]+H$\beta$ line emission in $z_{\rm{spec}}\sim6.3$ galaxies.

Along with this publication, we publicly release full reduced mosaics and a photometric catalog. Our website, \href{https:/jwst-uncover.github.io/megascience/}{jwst-uncover.github.io/megascience}, has links to all data products, and the associated Zenodo release is located at \href{https://zenodo.org/doi/10.5281/zenodo.8199802}{zenodo.org/doi/10.5281/zenodo.8199802}. 

Throughout this paper we assume a standard $\Lambda$CDM cosmology with $\Omega_m = 0.3$, $\Omega_\Lambda=0.7$, and $h=0.7$. Magnitudes are quoted in the AB system \citep{oke74}. All ID numbers given correspond to the DR3 UNCOVER+MegaScience catalogs released along with this paper.


\section{Observations}
\label{sec:observations}

\subsection{Survey Design \& Field}

The primary goal of the Medium Bands, Mega Science survey (``MegaScience"; PID 4111) is to obtain deep (2.3-4.6hr) imaging with the full suite of NIRCam medium-band filters over the same area as the UNCOVER survey \citep{bezanson22}. These medium-band data significantly increase the spectral resolution of existing broad-band observations, and allow us to perform a wealth of both spatially-integrated science (e.g., more accurate redshifts, stellar masses, and star formation rates; see Section~\ref{sec:integrated-science}) and spatially-resolved science (e.g., maps of line emission and physical properties; see Section~\ref{sec:resolved_science}). 

\begin{figure*}
    \centering
    \includegraphics[width=0.95\textwidth]{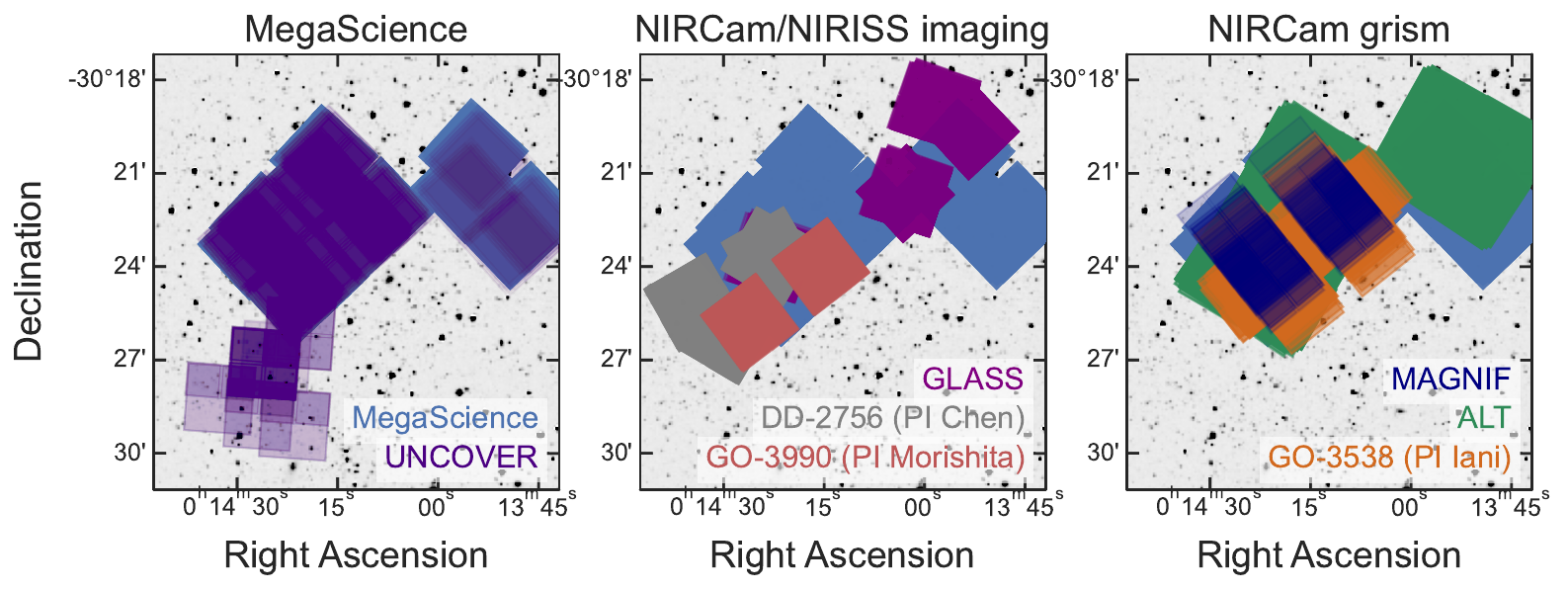}
    \caption{Schematic of the on-sky footprint of MegaScience and UNCOVER (left), other NIRCam and NIRISS imaging programs (center), and NIRCam grism programs (right); these programs are described in additional detail in the main text. The MegaScience footprint is shown in light blue in all three panels. A comparison with footprints of pre-JWST data, including HST and MUSE, is shown in \citet{bezanson22}. For the UNCOVER parallel observations in the southeast of the image, we only show pointings which are included in the reduced mosaics released along with this paper (declination $<30^\circ28$).}
    \label{fig:survey_footprint}
\end{figure*}

The Abell 2744 field hosts a remarkable compilation of public data, including {\it Hubble Space Telescope} imaging through the Hubble Frontier Fields program \citep{lotz17} and the BUFFALO survey \citep{steinhardt20}, deep MUSE spectroscopy \citep{mahler18}, and ALMA coverage \citep{fujimoto23}. Additionally, the field has been targeted by a number of early JWST programs including UNCOVER (\#2561; PIs Labb\'e \& Bezanson), the GLASS Early Release Science program \citep{treu22}, DDT program \#2756 \citep[PI: Chen;][]{chen_ddt}, a Cycle 2 pure parallel imaging program (\#3990, PI Morishita), parallel NIRCam imaging to Cycle 2 GLASS spectroscopy (\#3073; PI Castellano), as well as three Cycle 2  programs obtaining both NIRCam/grism and NIRCam/imaging (ALT, \#3516, PIs Matthee \& Naidu; MAGNIF, \# 2883, PI Sun; and \#3538, PI Iani). Footprints of pre-JWST data in Abell 2744 are shown in \citet{bezanson22}, while footprints of JWST data including the MegaScience survey are shown in Figure~\ref{fig:survey_footprint}. The vast majority of these JWST programs, including MegaScience, were taken with no proprietary time. All told, Abell 2744 has public data in both broad- and medium-band NIRCam imaging, NIRCam grism observations in multiple filters, and NIRSpec prism spectroscopy adding up to over 300hr of community investment in Cycle 1 \& 2 alone. This wealth of data means that Abell 2744 is rapidly emerging as one of the preeminent public extragalactic JWST deep fields.

\citet{furtak23} presents a state-of-the-art lensing model for Abell 2744 based off of the deep UNCOVER imaging. Out of the total $\sim29.2$ arcmin$^2$ area of our primary NIRCam pointing, 21.0 arcmin$^2$ (72\%) has magnification $\mu>1.5$, 10.5 arcmin$^2$ (36\%) has $\mu>2$, 2.3 arcmin$^2$ (8\%) has $\mu>5$, and 1.1 arcmin$^2$ (4\%) has $\mu > 10$; nearly every source in the footprint has $\mu\gtrsim1.3$\footnote{exact values calculated for $z=8$}. This lensing boosts our effective depth, allowing us to detect intrinsically fainter sources than a field survey with the same integration time. While the majority of moderately-magnified sources in the field are not significantly morphologically distorted, the small regions of high magnification near the cluster cores boast very highly magnified objects, allowing us to improve the effective spatial resolution by factors of up to $\sim5-10$.

\begin{figure*}[ht]
	\centering
	\includegraphics[width=0.95\textwidth]{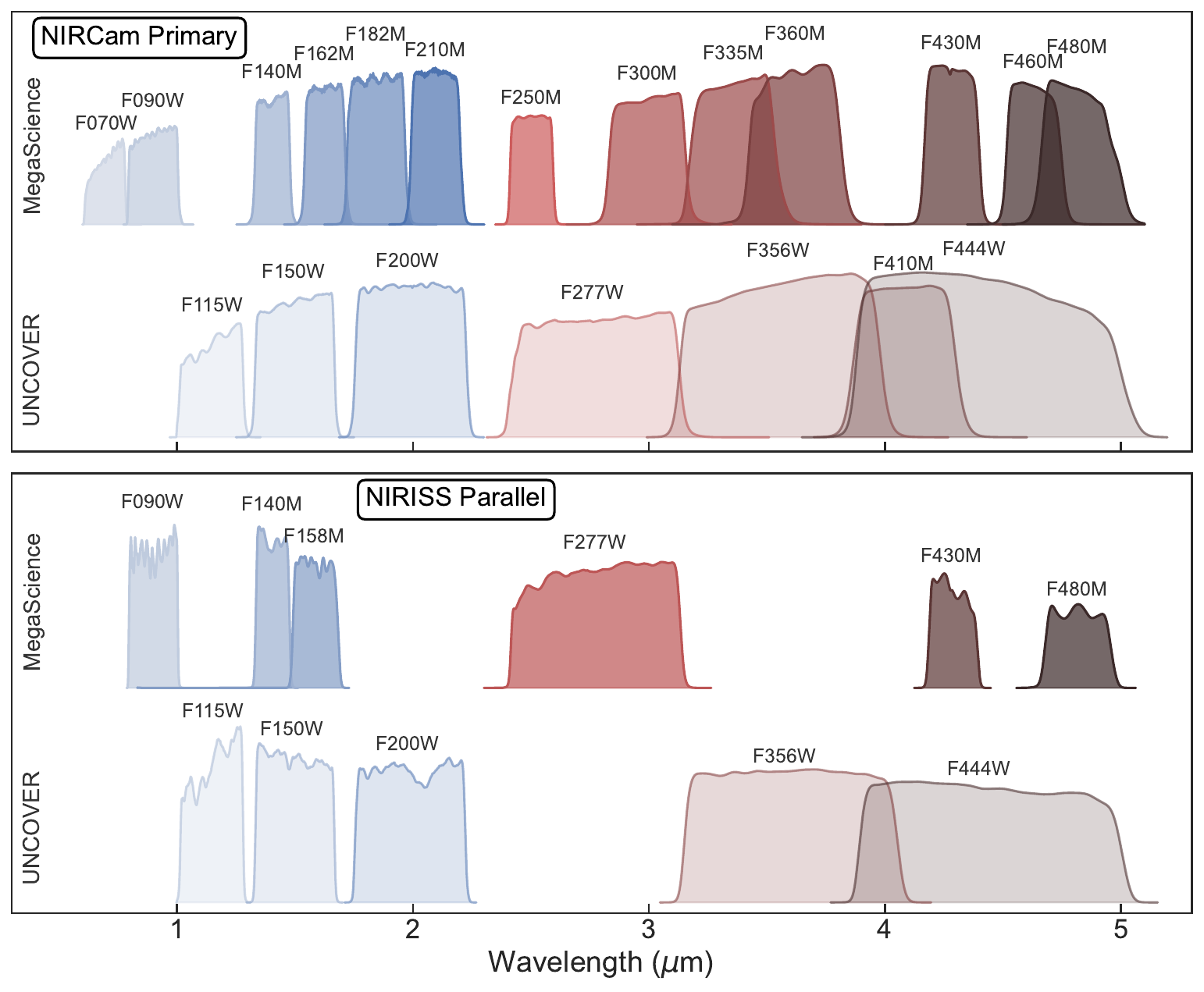}
	\caption{Existing photometric coverage in both the NIRCam primary and NIRISS parallel fields, along with new imaging from this program. MegaScience ``completes" the imaging in the primary field, with deep images in all 20 of NIRCam's broad- and medium-band filters. The addition of two broad-band and four medium-band filters in the NIRISS parallel field will allow for robust SED modeling. }
    \label{fig:filters}
\end{figure*}

As shown in Figure~\ref{fig:filters}, the primary NIRCam pointing of MegaScience targets all medium- and broad-band filters not observed as part of UNCOVER: eleven medium-band filters (F140M, F162M, F182M, F210M, F250M, F300M, F335M, F360M, F430M, F460M, and F480M; F410M was observed by UNCOVER) as well as two blue broad-band filters F070W and F090W. These blue broad-band filters can be used as dropout filters to select high-redshift candidates and are observed ``for free" during observations of NIRCam's numerous long-wavelength medium-band filters. The NIRISS parallel pointing fills in the existing broad-band UNCOVER photometric coverage with F090W and F277W, and additionally adds four medium-band filters F140M, F158M, F430M, and F480M. 

MegaScience was originally scheduled for four visits on November 5, 11, and 12, 2023. The first visit on November 5th failed due to guide star acquisition, and was rescheduled for November 20th with WOPR 88967. In order to obtain a suitable guide star and schedule the pointing in 2023, this rescheduled visit is slightly mis-aligned with the rest of the mosaic as shown in Figure~\ref{fig:survey_footprint}, resulting in a slight rotation relative to the existing UNCOVER data. 

\subsection{Data Reduction} 

The MegaScience data are reduced following similar methods as described in \citet{bezanson22}. All public NIRCam and NIRISS exposures in the Abell 2744 field are calibrated with the `jwst\_0995.pmap' calibration set after retrieval from MAST. This includes the MegaScience data discussed here (GO-4111, PI Suess), as well as imaging from UNCOVER \citep{bezanson22}, ERS-GLASS \citep{treu22}, DD-2756 (PI: Chen), MAGNIF (GO-2883, PI: Sun), GO-3538 (PI: Iani), and ALT (GO-3516; Naidu \& Matthee in prep.). Images are aligned and co-added with the \texttt{Grizli} pipeline (G. Brammer in prep., version 1.9.13.dev26), which uses the updated \texttt{snowblind} code{\footnote{https://github.com/mpi-astronomy/snowblind}} for masking NIRCam and NIRISS snowballs and an updated NIRCam bad pixel mask. These images correspond to the version 7.2 mosaic release hosted and described in additional detail on the DAWN JWST Archive (DJA; \url{https://dawn-cph.github.io/dja/imaging/v7/}).

As in UNCOVER, the brightest cluster galaxies (bCG) and associated intracluster light (ICL) are modelled and subtracted following \citet{shipley18} and described in detail in \citet{weaver24}. A subsequent background subtraction is also applied, using the same methodologies as the broad-band UNCOVER images. Table~\ref{tab:depths} presents the depths of the final, bCG/ICL-subtracted mosaics.

\begin{table}[ht]
\begin{threeparttable}[ht]
\caption{Measured MegaScience imaging depths}
\begin{tabular}{cccc}
\hline \hline
{Instrument}   & {Filter} & {Exposure} & {5$\sigma$ depth} \\ \hline
NIRCam  & F070W      &   2.3hr\tnote{1}       & 29.60     \\ 
(primary) & F090W & 4.6hr\tnote{1}  & 30.19 \\
& F140M &  2.3hr & 28.69 \\
& F162M & 2.3hr  & 28.67 \\
& F182M & 2.3hr  & 29.06 \\
& F210M & 2.3hr  & 29.02 \\
& F250M & 2.3hr  & 28.20 \\
& F300M & 2.3hr  & 28.66 \\
& F335M & 2.3hr  & 28.67 \\
& F360M & 2.3hr  & 28.62 \\
& F430M & 2.3hr  & 27.72 \\
& F460M & 2.3hr  & 27.44 \\
& F480M & 2.3hr  & 27.40 \\ \hline
NIRISS & F090W & 4.3hr & 28.93 \\
(parallel) & F140M & 2.1hr & 28.68 \\
& F158M &  2.1hr & 27.77 \\
& F277W &  2.1hr & 28.85 \\
& F430M & 2.1hr  & 27.07 \\
& F480M & 2.1hr  & 26.94 \\ \hline
\end{tabular}
\begin{tablenotes}
\item NOTE: Imaging depths are calculated in the same manner as \citet{bezanson22}, and correspond to the two-visit depths regions of the mosaic; depths are calculated using 0.08" radius apertures in the SW bands and 0.16" radius apertures in the LW bands and corrected to total assuming point sources.
\item[1] These exposure times represent only the MegaScience depth, but the images are significantly deeper due to imaging from ALT (R. P. Naidu \& J. Matthee in prep): median exposure times from ALT are 3.4hr in F090W and 4.3hr in F070W. 
\end{tablenotes}
\label{tab:depths}
\end{threeparttable}
\end{table}

\subsection{Source Detection and Photometry}

Multi-wavelength photometric catalogs are constructed with \texttt{aperpy}\footnote{\noindent \texttt{aperpy} is available though Github (\url{https://github.com/astrowhit/aperpy}) and Zenodo (\url{https://doi.org/10.5281/zenodo.8280270}).\label{foot:aperpy}} broadly following methods used to create the UNCOVER catalog \citep{weaver24}. Because the depth and area of long-wavelength broad-band imaging in Abell~2744 has increased significantly over the course of Cycles 1 \& 2, we construct a new 56.2\,arcmin$^2$ detection image as a noise-weighted stack of F277W, F356W, and F444W after bright cluster galaxies (bCGs) are removed. Using the same \texttt{SEP} \citep{barbary16} configuration as \citet{weaver24}, we detect a total of 74,020 objects. This is a $\sim20$\% increase compared to the DR2 UNCOVER catalog.

\begin{figure}
    \centering
    \includegraphics[width=\hsize]{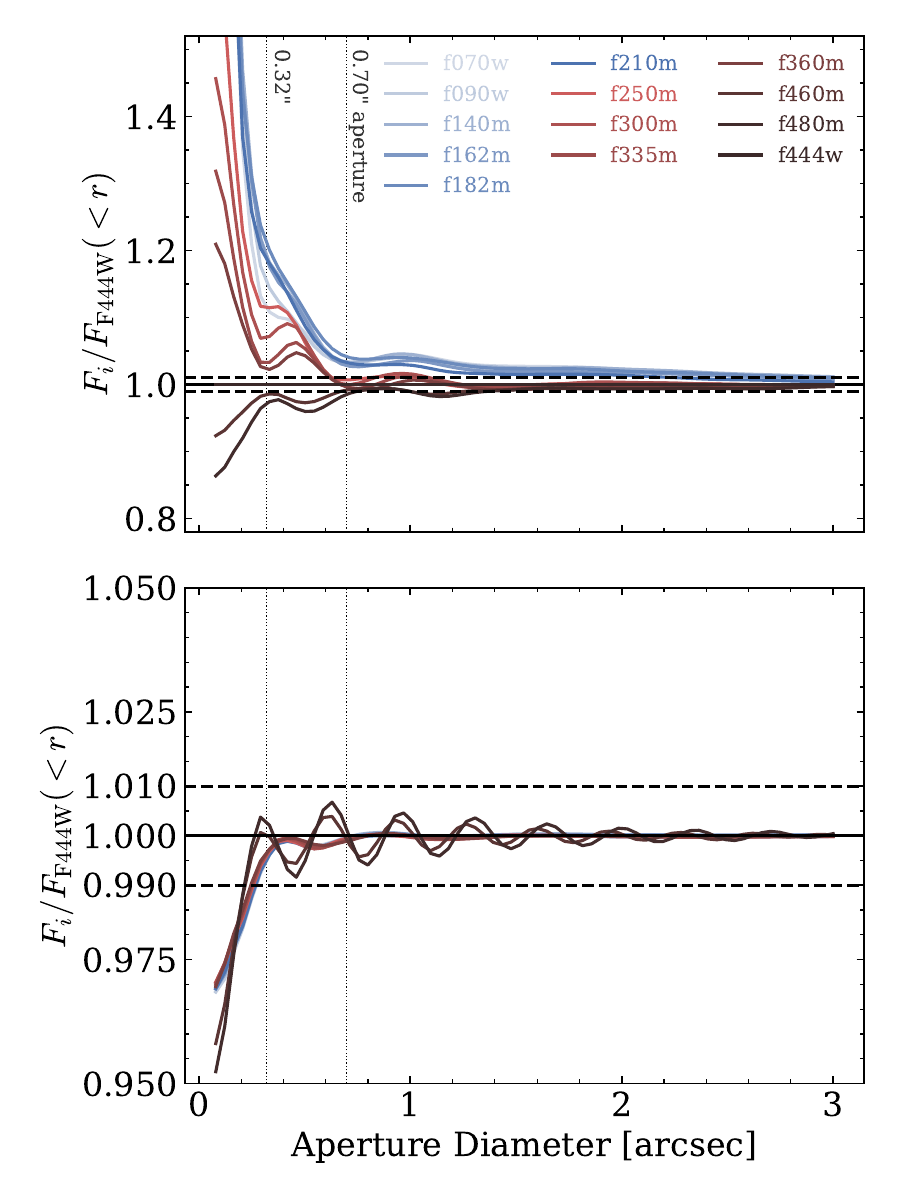}
    \caption{PSF growth curves for each filter added to UNCOVER before (top) and after (bottom) matching to F444W. After matching, all filters have deviations below the 1\% level at the smallest aperture diameter used (0.32\arcsec{}). Growth curves are shown relative to the F444W growth curve; a value of 1 indicates perfect matching with F444W. Dashed lines indicate the $\pm1$\% deviations from exact matching (solid black line). Dotted lines indicate the location of 0.32\arcsec{} and 0.70\arcsec{} aperture diameters.}
    \label{fig:psf_match}
\end{figure}

We measure photometry on PSF-matched images to account for the large variation in the PSF from different filters and instruments. While F480M has the broadest PSF in our survey, we choose to match to F444W for three reasons: (1) the F444W imaging is considerably deeper (by $\sim1.8$mag), allowing us to construct a robust empirical PSF; (2) F480M has a significantly more complex PSF than F444W, which would introduce additional complexity in the homogenization process; and (3) matching to F444W allows us to perform 1:1 comparisons with most other early JWST surveys, which typically match to F444W. We construct matching kernels to F444W using \texttt{Pypher} \citep{boucaud16}. 
Figure~\ref{fig:psf_match} shows the performance of our matching kernels for the filters observed in MegaScience, finding good agreement with the target F444W PSF below 0.1\% for most bands. The worst performance is from the four 4\,$\mu$m kernels (F410M, F430M, F460M, F480M) due to the rapid changes in the PSF structure relative to F444W, but even these long-wavelength PSFs agree well within 1\%.

Fluxes are extracted in four apertures of 0.32", 0.48", 0.70", 1.00", and 1.40" diameter and corrected to total fluxes by scaling to `AUTO' fluxes extracted from a PSF-matched version of our LW detection image in elliptical Kron apertures. We perform an additional correction for light outside the Kron radius \citep{kron80}
, following the methods of \citet{weaver24} \citep[see also][]{whitaker11,skelton14,whitaker19a}. Flux errors are estimated directly from the images using 10,000 apertures placed randomly on background regions. 
We construct a \texttt{USE\_PHOT} flag following similar methods as \citet{weaver24}; this flag is intended to remove both stars and imaging artifacts. Similar to the UNCOVER catalog, the vast majority (91\%) of objects have \texttt{USE\_PHOT = 1}. For ease-of-use, we provide a `SUPER' catalog wherein photometry is extracted in the optimal aperture for each object, taking extra care to avoid failures in crowded regions of these deep images. In practice, lower redshift galaxies with extended light profiles will default to larger apertures unless they are in densely crowded regions where it is better to use a smaller aperture to avoid contamination. 

As a check, we compare our newly extracted photometry to UNCOVER DR2 in all common, pre-MegaScience bands, finding zero bias at all magnitudes and a scatter commensurate with expected flux uncertainties. In other words: the photometry in our catalog agrees well with the previous public release of the UNCOVER catalog in the broad bands, despite the new segmentation map. As discussed in \citeauthor{weaver24}, it is difficult to assess photometric reliability (especially for the new medium bands) without comparing to an independent catalog. \citet{weaver24} perform a sanity check by comparing the catalog fluxes to the predicted photometry from \texttt{eazy}'s best-fit model. We perform a similar test using the MegaScience catalog, and, like UNCOVER, we find good agreement from all bands. We refer the reader to the documentation accompanying these catalogs for up-to-date details as to its contents, column names, and conventions.

\subsection{Photometric Redshifts}
\label{subsec:photoz}

One of the motivations behind this survey is to significantly improve photometric redshift estimates ($z_{\rm phot}$) and related physical parameters. As with the original UNCOVER survey, our team will provide two sets of redshift and stellar population catalogs: $z_{\rm phot}$ estimates fit using $\texttt{eazy-py}$ \citep{brammer08}, and a stellar population catalog fit using \texttt{Prospector} \citep{johnson21}, following the methods described in \citet{wang24}. 

We estimate $z_{\rm phot}$ by fitting all HST and JWST photometry in the default `SUPER' catalog with \texttt{eazy-py} \citep{brammer08}, following the same methods as \citet{weaver24}. We provide $z_{\rm phot}$ estimates run with the \texttt{SFHz\_BLUE\_AGN} template set, which includes redshift-dependent priors to ensure that galaxies cannot be fit with ages older than the age of the Universe. These templates also include an AGN torus template from \citep{killi23}. Priors on magnitude and $\beta$ slope are switched off. We append the estimated photometric redshifts and select rest-frame fluxes to our photometric catalogs. The photometric catalog, $z_{\rm{phot}}$ estimates, and reduced bCG-subtracted images are publicly available on our website (\href{https:/jwst-uncover.github.io/megascience/}{jwst-uncover.github.io/megascience}).

For the \texttt{Prospector} catalog, we fit the same HST and JWST photometry, but now using the methods and physical model described in \citet{wang24}. The catalog of best-fit redshifts is released along with this paper; a catalog of best-fit stellar population parameters (masses, star formation rates, etc) will be presented in B. Wang et al. (in prep.) along with a detailed description of the improvements in physical parameters using the medium-band photometry.

The performance of both sets of photometric redshifts is discussed in more detail in Section~\ref{sec:integrated-science} and Figure~\ref{fig:zphot_zspec}.


\section{Science Objectives}
\label{sec:science}

\subsection{Spatially-integrated science: improving redshifts, stellar masses, and star formation rates}
\label{sec:integrated-science}
The MegaScience dataset samples the SEDs of all galaxies in the field at nearly the highest spectral resolution possible with NIRCam imaging\footnote{NIRCam also includes seven narrow-band filters, which are often impractical for deep-field galaxy science due to the faint nature of the sources.}. This spectral resolution allows us to improve the accuracy of recovered photometric redshifts and stellar population properties while breaking degeneracies introduced by fitting broad-band photometry. In this section, we demonstrate the improvement MegaScience is able to make on the spectral energy distributions of galaxies compared to UNCOVER broad-band photometry. 

\begin{figure*}
    \centering
    \includegraphics[width=.95\textwidth]{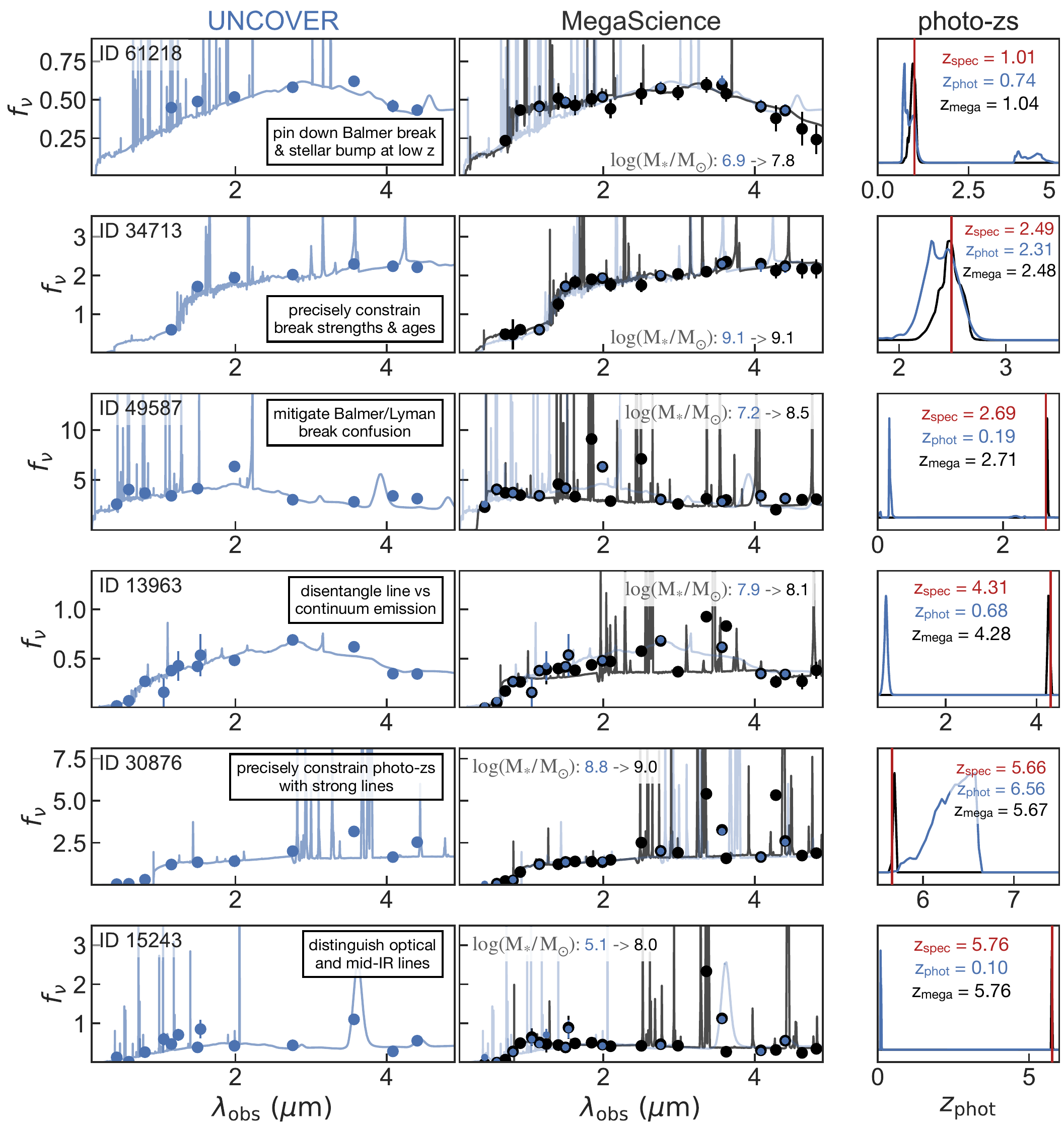}
    \caption{Example SEDs from UNCOVER (left) vs MegaScience (center), as well as \texttt{eazy-py} photometric redshift distributions (right). SEDs are shown in $f_\nu$ (in units of $10^{30}\rm{erg\ s}^{-1}\rm{cm}^{-2}\rm{Hz}^{-1}$). Each object is spectroscopically confirmed by UNCOVER (top four objects; Price et al. in prep.) or ALT (bottom two objects; Naidu \& Matthee et al. in prep). Objects are ordered by increasing redshift; each is chosen to demonstrate one ``failure mode" of SED-fitting that is solved by the increased photometric coverage of MegaScience. Each example is discussed in additional detail in Section~\ref{sec:integrated-science}.}
    \label{fig:seds}
\end{figure*}

Figure~\ref{fig:seds} shows the photometry and \texttt{eazy-py} fits to six example galaxies, all of which have spectroscopic redshifts from either UNCOVER (IDs 61218, 34713, 49587, and 13963; S. Price et al. in prep) or ALT (IDs 30876, 15243; J. Matthee \& R. Naidu et al. in prep). The left column shows the fit to UNCOVER data alone (blue); the central column shows the new MegaScience photometry and the corresponding \texttt{eazy-py} fit (black); the right column shows the photometric redshift posteriors of both the UNCOVER and MegaScience fits compared with the spectroscopic redshift (red). Each galaxy was chosen to represent a way that the medium-band data resolves a common ``failure mode" of broad-band SED fitting.

ID 61218 shows a star-forming galaxy at $z\sim1$. In this redshift regime, the MegaScience data is able to both pin down the location of the Balmer break (thanks to F070W \& F090W data, especially important in the portions of the field lacking HST coverage), and identify the 1.6$\mu$m stellar bump with the long-wavelength medium-bands. Together, these features allow for precise photometric redshift recovery and effectively rule out high-redshift solutions that are allowed by the UNCOVER fit alone. We expect the 1.6$\mu$m stellar bump to also help identify cluster members in Abell 2744 itself at $z\sim0.33$. This is especially important when pushing down the mass function towards the lowest mass cluster members, where the 1.6 micron bump is effectively the only prominent feature identifiable in the absence of adequate coverage of the Balmer break (observable at $<0.5$ micron).

ID 34713 shows a galaxy at cosmic noon ($z\sim2.5$) with relatively low SFR~$\sim 1-10M_\odot$yr$^{-1}$, potentially a massive galaxy caught early in the quenching process. For this class of object, the MegaScience data is able to (a) confirm low rest-UV fluxes with new 0.7-1.5$\mu$m data, indicating low ongoing SFRs, (b) precisely constrain the observed wavelength of the Balmer break, and (c) constrain the shape of the spectrum redward of the break. The location and shape of the Balmer break are particularly useful for characterizing the ages of quenching and quiescent galaxies \citep[e.g.,][]{carnall19,wild20,suess22}.

ID 49587 shows an example where where the \texttt{eazy-py} fit using UNCOVER data alone prefers a Balmer break, but the MegaScience fit reveals the (accurate) Lyman break solution. The degeneracy between Balmer and Lyman breaks is one of the most common causes of low-redshift interlopers in high-redshift photometric selections. In large part, the MegaScience data is able to break this degeneracy by identifying strong line emission boosting in several of the medium-band filters; no lines are predicted in these locations for the Balmer-break model. 

ID 13963 shows another classic photometric redshift mixup: what is fit with a rising red continuum in the UNCOVER data is revealed to be boosting by line emission in the MegaScience data. The true continuum level can be seen in the F300M filter, which is one of the few line-free filters in the entire SED $<4\mu$m. JWST medium bands have been shown to be highly effective in disentangling these line vs continuum solutions for large samples of distant galaxies \citep[e.g.,][]{laporte23,desprez23,sarrouh24}.

ID 30876 demonstrates the improvement in both accuracy and precision of photometric redshifts enabled by medium-band observations of sources with strong line emission. This galaxy, confirmed at $z=5.66$ with ALT grism observations (J. Matthee \& R. Naidu et al. in prep), had a relatively broad photometric redshift posterior with UNCOVER alone ($\delta z\sim0.43$): the fit was unable to precisely constrain the location of the Lyman and Balmer breaks, and it was unclear where exactly the H$\beta$+[OIII] lines were. The MegaScience data nails down the locations of the Lyman break, the Balmer break, H$\beta$+[OIII], and H$\alpha$, resulting in an accurate photometric redshift with an uncertainty of just $\sim$0.02. 

ID 15243 demonstrates a uniquely-JWST photometric redshift mixup: the H$\beta$+[OIII] and H$\alpha$ lines at $z\sim5-7$ are mistaken for the 3.3 PAH feature and the Br-$\alpha$ line at low redshift. This degeneracy appears to be relatively common in our initial \texttt{eazy-py} fits; however, we note that the low-redshift fits also have very low inferred stellar masses of $10^{4-6}M_\odot$ -- in essence, a high-redshift star-forming galaxy is mistaken for a low-redshift globular cluster. Due to the low inferred mass of this low-redshift solution, this degeneracy is unlikely to be common using codes such as \texttt{Prospector} which enforce a lower limit on the allowed stellar mass. We note that these very low-mass solutions present an opportunity to implemented a linked mass-dust prior to further aid disentangling degenerate solutions \citep[e.g.,][]{nagaraj22,alsing24}.

\begin{figure*}
    \centering
    \includegraphics[width=.85\textwidth]{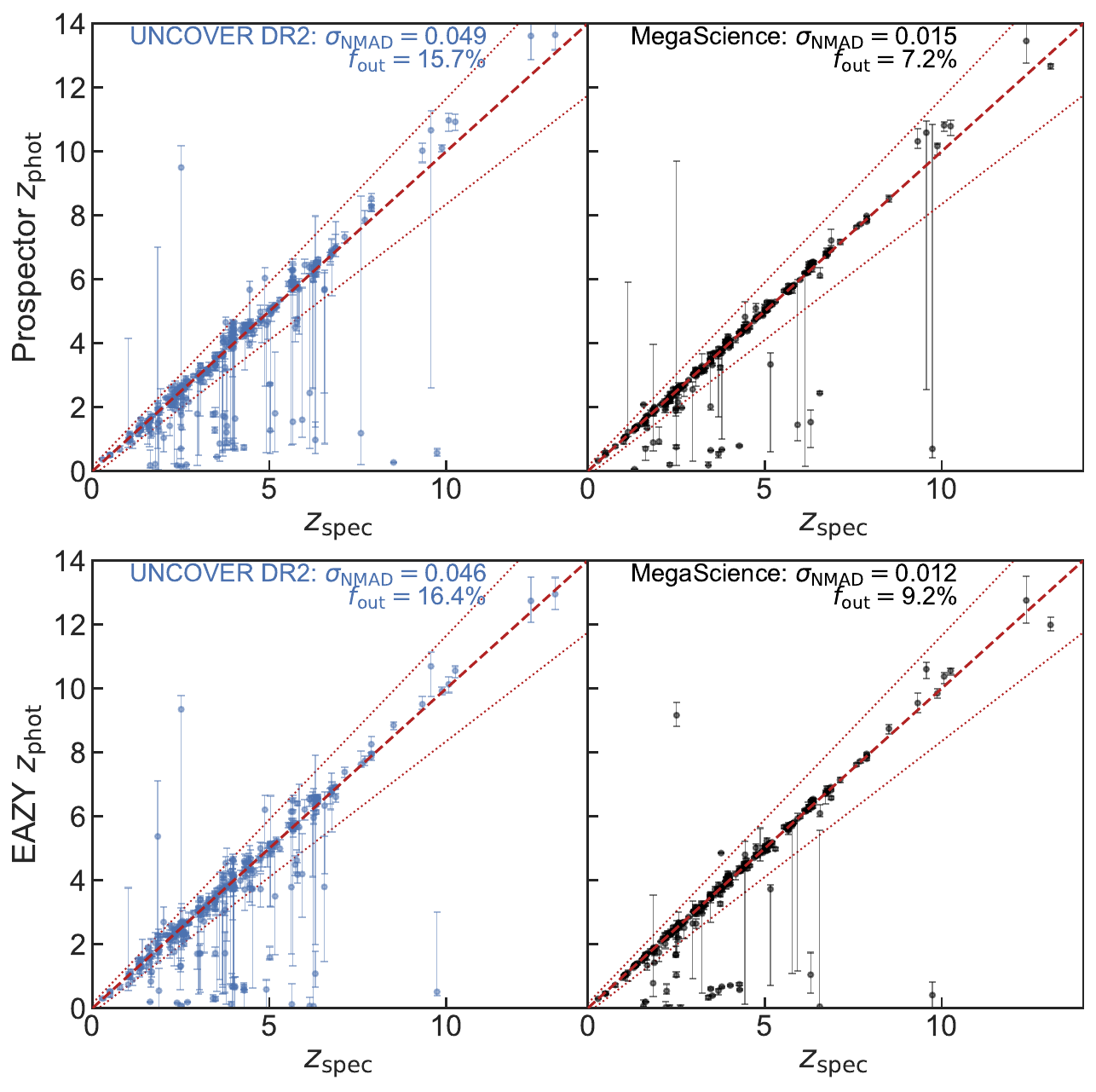}
    \caption{Comparison of photometric redshifts with a sample of $\sim300$ high-quality spectroscopic redshifts from the UNCOVER survey (S. Price et al., in prep). The top row shows our \texttt{Prospector} redshifts \citep[for details, see][]{wang24}, and the bottom row shows \texttt{eazy-py} redshifts based on photometry from the `SUPER' catalog \citep[for details, see][]{weaver24}. The left column shows redshifts from the public DR2 UNCOVER data release, while the right column shows MegaScience redshifts from the DR3 public data products accompanying this paper. MegaScience shows nearly a factor of {\it three} reduction in $\sigma_{\rm{NMAD}}$, and close to a factor of two reduction in catastrophic outliers (as in \citealt{weaver24}, defined as objects with $\Delta z / (1+z_{\rm{spec}})\ge 0.15$; shown by dotted red lines). \texttt{Prospector} and \texttt{eazy-py} show similar performance.}
    \label{fig:zphot_zspec}
\end{figure*}

The MegaScience improvements on these individual SED fits adds up to a substantial improvement in overall photometric redshift precision. In Figure~\ref{fig:zphot_zspec}, we show the $z_{\rm phot}$ performance from both our \texttt{Prospector} (top) and \texttt{eazy} (bottom) spectrophotometric redshift catalogs relative to a sample of $\sim300$ spectroscopic redshifts from the UNCOVER survey (S. Price et al. in prep.). We quantify this performance using the normalized median absolute deviation, $\sigma_{\rm{NMAD}}$, as defined in \citet{brammer08} and \citet{weaver24}, and the catastrophic outlier fraction, defined as the fraction of objects with $|z_{\rm{phot}}-z_{\rm{spec}}| / (1+z_{\rm{spec}}) \ge 0.15$. 
We find that MegaScience cuts the scatter $\sigma_{\rm{NMAD}}$ by a factor of {\it three}, and nearly halves the rate of catastrophic outliers. While the set of spectroscopically-confirmed galaxies from UNCOVER is a small and biased subset of the full galaxy population, this improvement in scatter and catastrophic outlier fraction is suggestive of a dramatic improvement in our catalog's overall photometric redshift precision and accuracy. Future comparisons, especially with grism redshifts, will help verify the $\sim1.6$\% scatter quoted in this paper.

In sum: the increased spectral resolution of the MegaScience observations are able to break common degeneracies in photometric redshift fitting, as well as improve the accuracy and precision of recovered physical quantities including photometric redshifts by factors of 2-3. Combining this data with the large set of ancillary data available in Abell 2744 -- MUSE, HST, ALMA, and now JWST/NIRCam-grism and NIRSpec -- will further improve our inferences about the physical properties of these distant objects.

\subsection{Spatially-resolved science: mapping galaxy growth across cosmic history}
\label{sec:resolved_science}

\begin{figure*}[ht]
	\centering
	\includegraphics[width=0.9\textwidth]{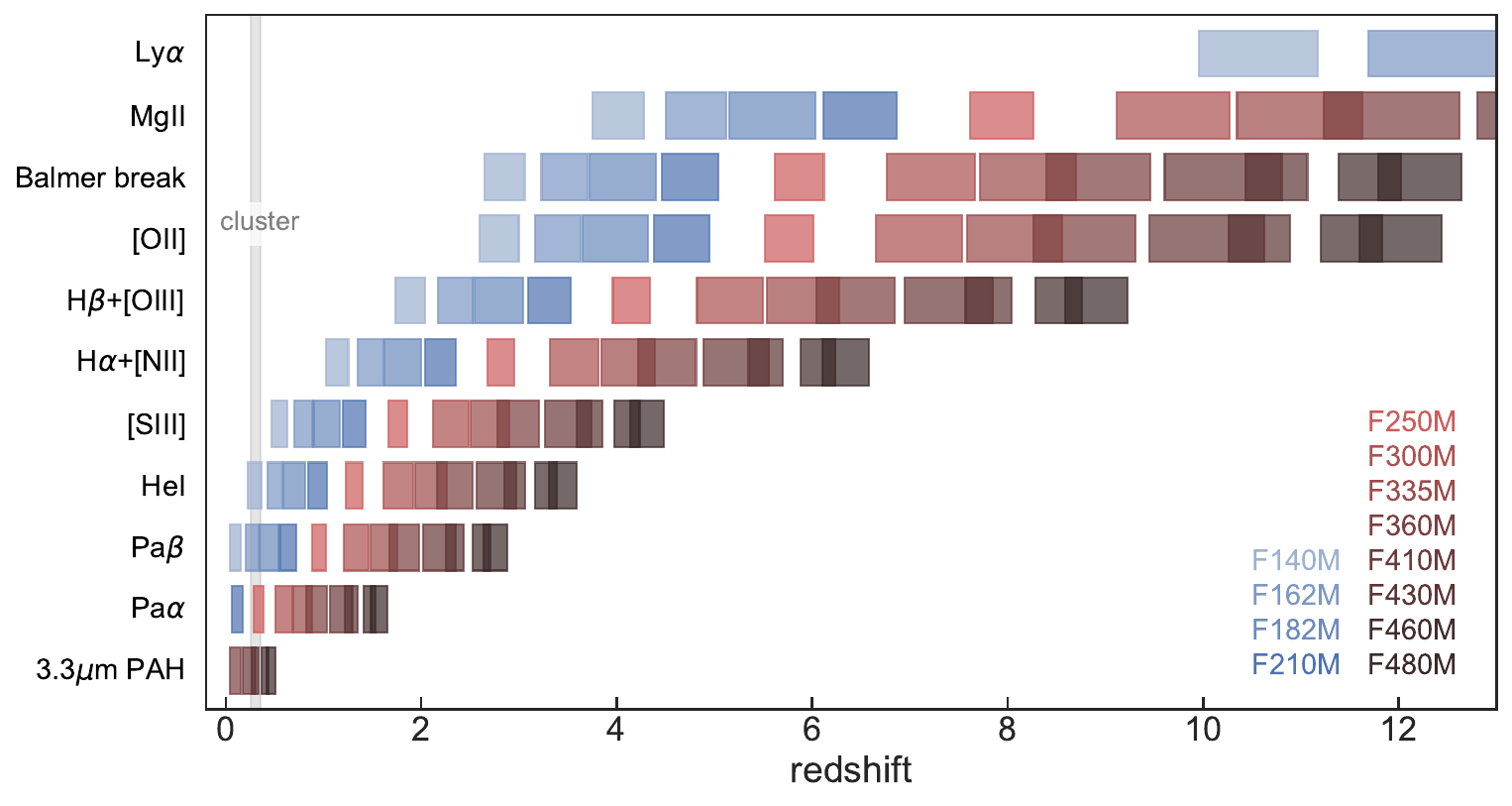}
	\caption{Spectral features probed by our primary medium-band observations as a function of redshift.}
        \label{fig:lines}
\end{figure*}

The MegaScience dataset not only provides high spectral resolution photometry: the high {spatial} resolution provided by NIRCam allows us to map the spatially-resolved properties of galaxies across $>10$~Gyr of cosmic history. Unlike grism data, where best practices often involve filtering out continuum emission to reduce contamination (e.g., \citealt{matthee23_eiger}), medium-band observations are sensitive to both continuum and line emission. This means that our data enable accurate 2D maps of both emission lines and, via spatially-resolved SED modeling, stellar population properties such as stellar mass and age. 
Figure~\ref{fig:lines} shows a range of strong spectral features that are accessible with each NIRCam medium band across redshift, from the cluster itself through cosmic dawn. Our survey strategy (``ALL the medium bands!") means that at most redshifts we cover multiple spectral features simultaneously. 

The remainder of this section describes the spatially resolved science achievable at various redshift epochs and provides example medium-band images of galaxies in our survey. We note that all images shown in this section have been gridded onto a common pixel scale of $0\farcs04$, but images have {\it not} been PSF-matched.

\subsubsection{The Low-redshift Universe: Star Formation, Stripping, and Dust Emission in Abell 2744}

As shown in Figure~\ref{fig:lines}, our imaging directly probes the 3.3$\mu$m PAH feature, Pa-$\alpha$, Pa-$\beta$, and HeI ($10830\AA$) + Pa-$\gamma$ in the Abell 2744 cluster itself. While Pa-$\beta$ and HeI+Pa-$\gamma$ are unlikely to have high enough equivalent widths to observe via medium-band excesses, given our depth in the F250W filter (Table~\ref{tab:depths}) we expect to observe Pa-$\alpha$ for any regions within cluster galaxies with SFRs~$\gtrsim0.5M_\odot$yr$^{-1}$.  

\begin{figure*}
    \centering
    \includegraphics[width=.95\textwidth]{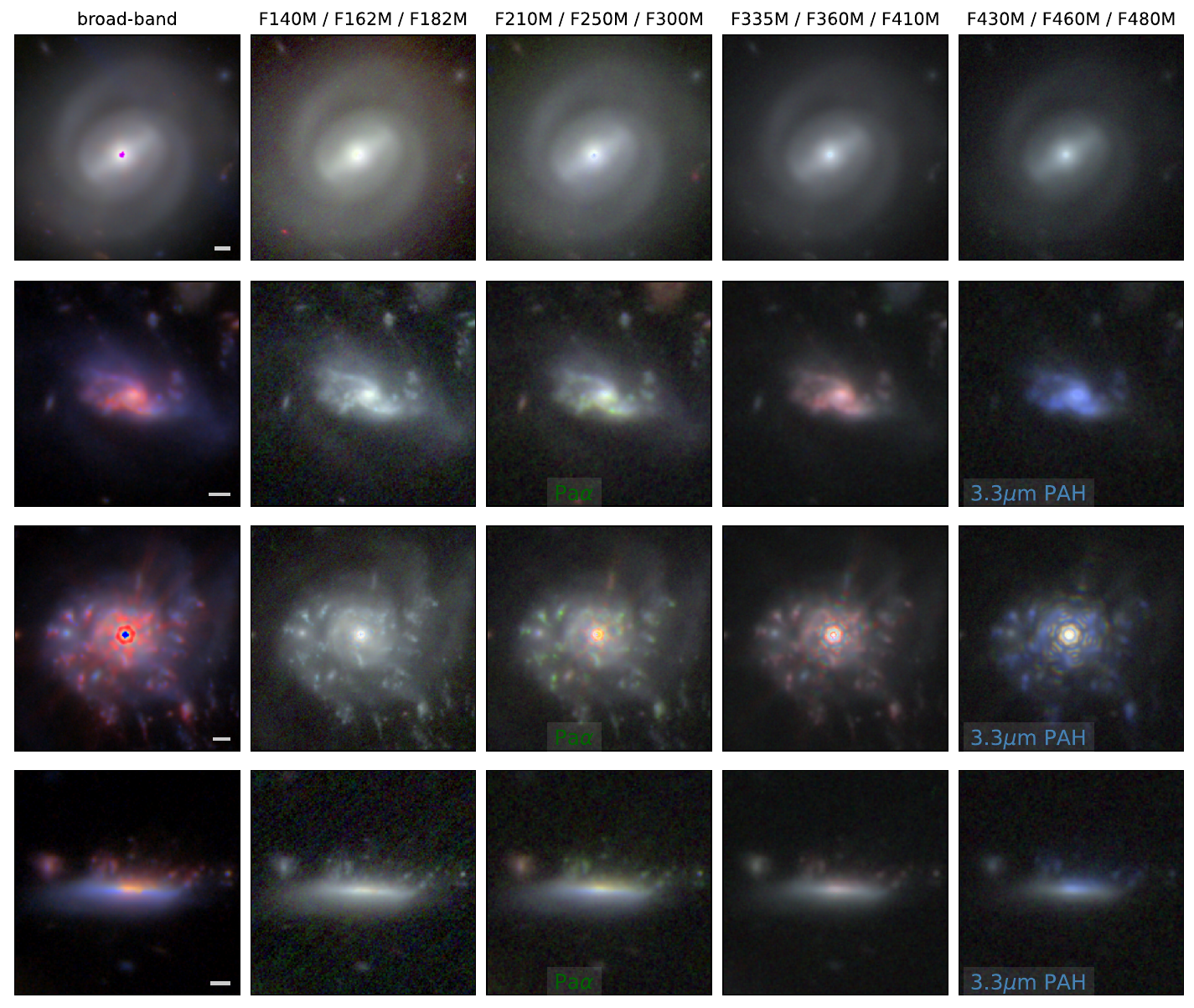}
    \caption{Examples of four galaxies in the Abell 2744 cluster itself; each object uses the same arcsinh color stretch. While most cluster galaxies have low ongoing SFRs and relatively uniform colors (like the top object), the cluster also hosts several spectacular morphologically-disturbed galaxies with strong Pa$\alpha$ and 3.3$\mu$m PAH emission. The white scale bar shows 0\farcs5.}
    \label{fig:cutouts_cluster}
\end{figure*}

Figure~\ref{fig:cutouts_cluster} shows examples of four cluster galaxies; the leftmost column shows a broad-band image constructed using the F115W, F277W, and F444W filters from UNCOVER \citep{bezanson22}; the remaining four color images are constructed using all dozen MegaScience medium bands as listed in the column headings. Typical cluster galaxies have minimal ongoing specific star formation rates and resemble the top object: relatively uniform in color across all medium and broad bands. However, the cluster hosts several spectacular examples of galaxies experiencing large starbursts, potentially driven by ram-pressure stripping \citep[for examples at similar $z\sim0.3$, see][]{vulcani24}. In these three galaxies, both Pa-$\alpha$ (F250M) and the broad 3.3$\mu$m PAH feature (F410M \& F430M) are clearly visible by eye. This emission varies across the galaxy, indicating local regions of elevated star formation and PAH emission \citep[as discussed in detail in][]{spilker23}. Our observations allow us to map star formation and dust  in these active cluster galaxies, as well as search for localized star-forming regions in cluster galaxies that are globally quiescent.

\subsubsection{Cosmic Noon: $1\lesssim z\lesssim 3$}
At cosmic noon, our imaging is able to constrain multiple spectral features (especially Balmer lines) simultaneously; at $z\sim1$ we directly probe Pa$\alpha$, Pa$\beta$, and H$\alpha$+[NII], while at $z\sim3$ we directly probe Pa$\beta$, H$\alpha$+[NII], H$\beta$+[OIII], and the Balmer break. Careful modeling -- likely informed by the wealth of spectroscopic observations in Abell 2744 -- will be required to tease out the relative contributions of H$\beta$ and [OIII] \citep[e.g.,][]{kaasinen17}. 

\begin{figure*}
    \centering
    \includegraphics[width=.9\textwidth]{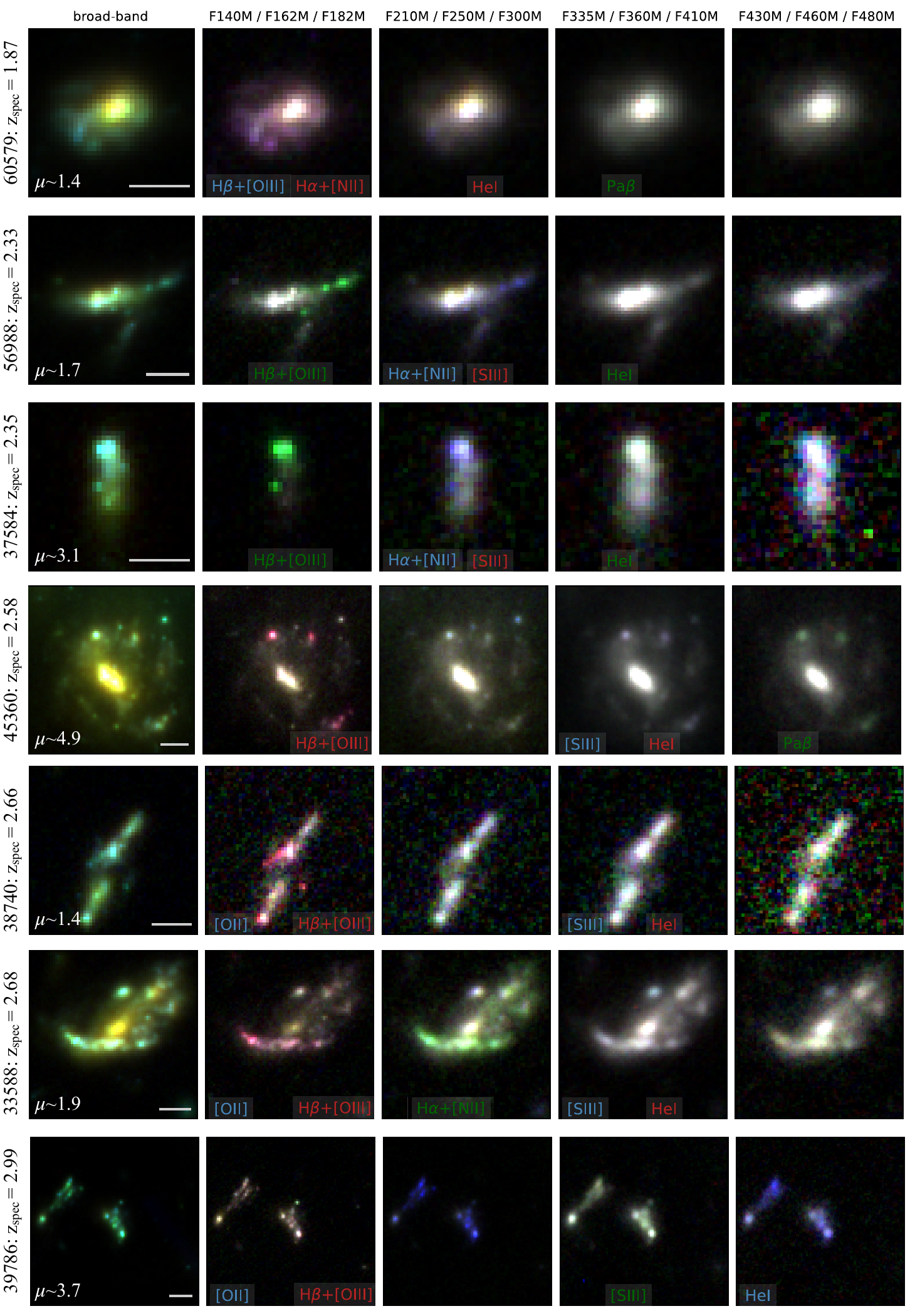}
    \caption{Examples of galaxies at cosmic noon, $1\lesssim z\lesssim3$. Color images are shown on a linear stretch; the white scale bar shows 0\farcs5, which corresponds to $\sim4$~kpc at these redshifts.}
    \label{fig:cutouts_noon}
\end{figure*}

Using these multiple line diagnostics, MegaScience is capable of simultaneously mapping star formation and dust attenuation within galaxies at cosmic noon. By combining these maps with structural fits to the galaxies, we plan to study how the relationship between galaxy star formation histories and structure builds up over time. Due to the high spatial resolution of NIRCam -- and aided by strong gravitational lensing in the cluster, see \citet{furtak23} -- it is also possible to study the clumpiness of both star formation and dust in massive galaxies at the height of cosmic star formation history \citep[e.g.,][]{lu24,ji23}. 

Figure~\ref{fig:cutouts_noon} shows example images of seven galaxies at $1\lesssim z \lesssim 3$. The leftmost column shows a broad-band image constructed using UNCOVER data; the other four columns show all NIRCam medium bands. The structures of these galaxy are extremely complex: even in seemingly-well-settled disks such as ID 56988, MegaScience observations reveal clumps of H$\alpha$ lighting up in the spiral arms. In many objects, star-forming clumps can be seen lighting up in multiple different strong emission lines including H$\beta$+[OIII], HeI+Pa$\gamma$, [SIII], and Pa$\beta$.

\subsubsection{Cosmic Morning: $3\lesssim z\lesssim6.5$}

At $3 \lesssim z\lesssim 6.5$, our data begin to probe the Balmer break and the [OII] and MgII lines, in addition to the H$\beta$+[OIII] and H$\alpha$+NII lines. At these redshifts, our data are able to map star formation and dust attenuation, identify AGN, and potentially identify and constrain some of the first quiescent galaxies \citep[e.g.,][]{carnall23,setton24}. Importantly, including a suite of (fewer) medium bands in the restframe optical has already been shown to work well to photometrically select quiescent galaxies during this early era, even in the absence of the restframe-J band data \citep[only probed by MIRI at these wavelengths;][]{alberts2024}. 

\begin{figure*}
    \centering
    \includegraphics[width=.9\textwidth]{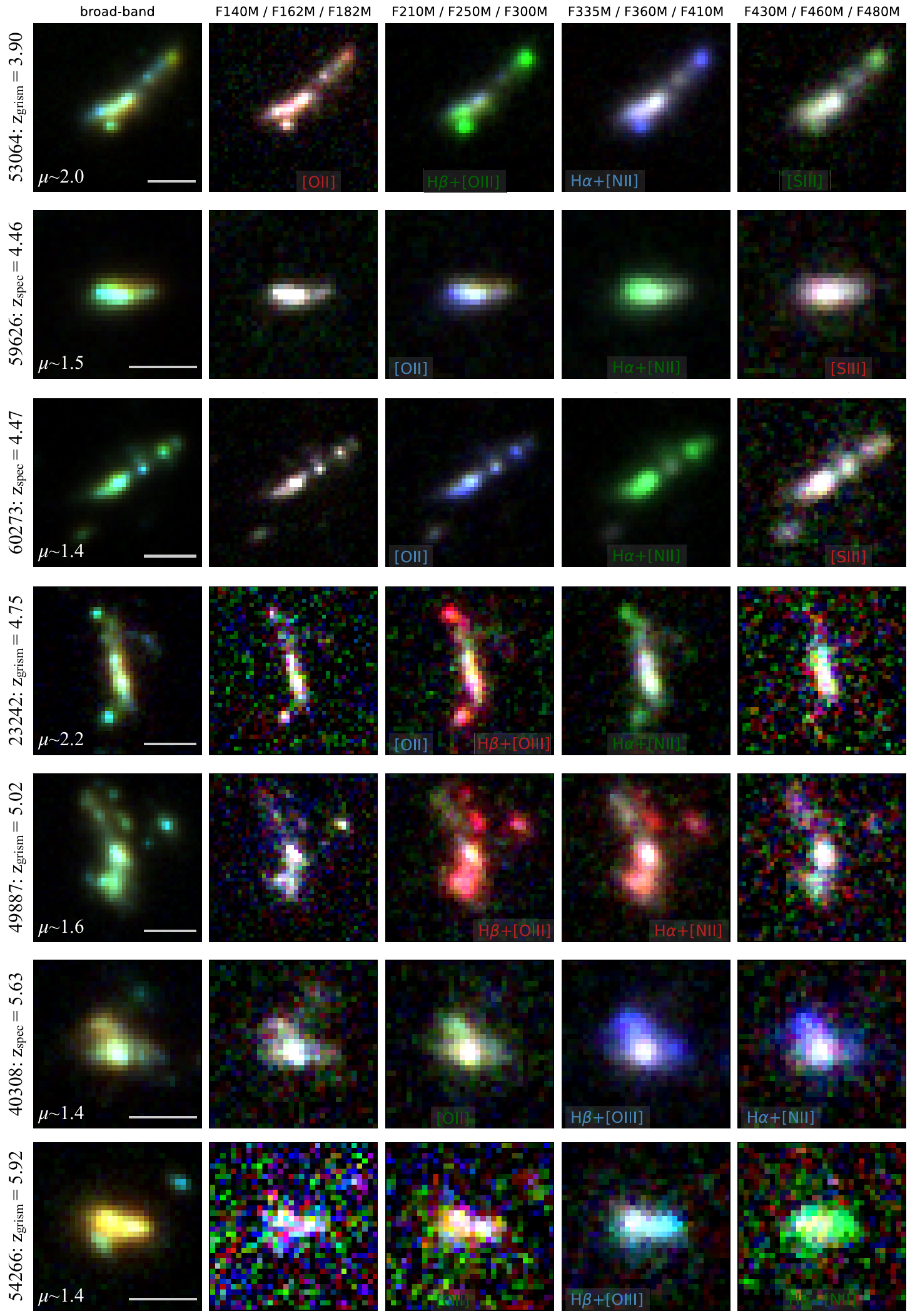}
    \caption{Example of galaxies at cosmic morning, $3\lesssim z\lesssim6$. Color images are shown on a linear stretch; the white scale bar shows 0\farcs5.}
    \label{fig:cutouts_elevensies}
\end{figure*}

Figure~\ref{fig:cutouts_elevensies} shows example images in this redshift range, following the same format as Figure~\ref{fig:cutouts_noon}. These galaxies are morphologically complex, often with multiple bright clumps. Significant line boosting is present in multiple emission lines, and in some cases emission line ratios show clear spatial variation even by eye. This highlights how the power of simultaneous coverage of H$\beta$+[OIII] and H$\alpha$+NII (provided by our medium bands up to $z<6.5$) at high spatial resolution can provide key insights into the drivers of these emission lines \citep[e.g. star formation bursts vs ionization][]{endsley2023} while controlling for biases present in integrated measurements \citep{gimenez-arteaga23}.

Our data has near-complete coverage of the H$\alpha$+NII line complex in this redshift range, with important cross-section at $z>3$ with Lyman-$\alpha$ probed by other facilities. Lyman-$\alpha$ emission (rest-frame 1216 $\AA$) experiences resonant scattering with neutral Hydrogen, making it a probe of the state of cosmic reionization \cite[e.g.][]{malhotra04}.  However, using Lyman-$\alpha$ as the single probe of reionization is complicated by the strong dependence of Lyman-$\alpha$ escape on internal galaxy properties (such as dust and young stellar populations) and observational effects \cite[such as bias in slit losses due to spatial offsets between rest-UV and Lyman-$\alpha$;][]{matthee16,hoag19}. These issues motivated panchromatic surveys of Lyman-$\alpha$ emitters at $z \lesssim 3$ that suggest that low metallicity stars with hard ionizing UV spectra are a key source of photons that can escape to ionize the IGM \cite[e.g.][]{steidel18}.

With MegaScience, we can directly observe H$\alpha$ instead of Lyman-$\alpha$ into the late epoch of reionization,  facilitating studies of the relative ionizing photon production efficiency of galaxies over a large range in mass and luminosity.  This can be coupled with the deep MUSE spectroscopy available in the field \citep{mahler18} covering Lyman-$\alpha$ out to $z=6.7$. Our full suite of filters expands the redshift range probed compared to earlier medium band studies \citep[e.g.][]{Simmonds2023,Simmonds2024}, enabling a more complete assessment of redshift evolution.  The strength of this approach is the ``untargeted’’ nature of both sets of observations, allowing us to study H$\alpha$ in NIRCam for cases where Lyman-$\alpha$ is not detected in the MUSE data, and vice-versa.  This will result in an unbiased and complete picture that is impossible with data sets that can only focus on a single observable.

\subsubsection{Reionization: $6.5\lesssim z<10$}

At this redshift range H$\alpha$+NII has redshifted out of the NIRCam wavelength coverage; however, our observations still probe the Balmer break, [OII], [NeIII] and H$\beta$+[OIII] (among bluer rest-UV probes). Strong emission features are common, and not all galaxies are pointlike: many still show complex, clumpy, extended morphologies (see Section~\ref{sec:demo}). While the inclusion of the two bluest broad-band filters F070W and F090W allows us to select high-redshift candidates with more confidence than in the UNCOVER data alone, the suite of 3-4$\mu$m medium-band filters drastically expands the astrophysical constraints on the ionizing radiation from Reionization-era sources \citep[as suggested by mock observations;][]{robertsborsani21}.  In particular, the simultaneous sampling of both [OIII]+H$\beta$ and the Balmer break 
improve constraints that discriminate between timescales of star formation history with other parameters that influence the line equivalent widths such as metallicity or ionizing escape fraction \citep[e.g.][]{endsley2023, laporte23}.


\subsubsection{Cosmic Dawn: $z\gtrsim10$}

Beyond $z>10$, NIRCam only probes the restframe UV of galaxies, yet MegaScience will reveal important diagnostics for the formation of the first stars and galaxies. These include the Lyman break, MgII, Balmer break if present, [OII] \citep[and neighboring NeIII, which JWST has revealed appears strong in early galaxies; e.g.][]{castellano24, RobertsBorsani24}, as well as exquisite sampling of the restframe UV continuum shape, $\beta$. Together, these spectral signatures can provide improved astrophysical insight into the ages and dust content of these earliest galaxies.

Robust identification of candidates, however, remains a challenge with broad band photometry, the workhorse of early JWST surveys in Cycle 1. The community has now gained enormous insight into the primary interloper population in the identification of galaxies at Cosmic Dawn. Namely, galaxies at $3<z<6$ can mimic high-redshift sources through some combination of balmer break, emission line boosting and dust \citep[e.g.][]{naidu22,zavala23,arrabal-haro23}. 
Better sampling of the dropout features improve the discriminating power between 
these scenarios, with sensitivity to the difference between a sharp Lyman-alpha break and the more gradual spectral shapes, and emission line boosting, of the primary lower-redshift interloper population
\citep[see e.g. Figure 2 in][]{eisenstein23_jof}. In principal, MegaScience provides medium-band based selections (or, upper limits to abundances in case of non-detection) from $z\gtrsim11$ (F140M-dropouts).

Already, medium bands are enabling improved interloper rejection, providing constraints on abundances and physical properties up to $z\sim14$ \citep{robertson24,willott24,bouwens23}. A critical improvement provided by MegaScience is the larger 3$\times$ area of coverage compared to some early medium band surveys (e.g. JEMS, JOF), since early galaxies are extremely rare and can be highly clustered. MegaScience nearly doubles the coverage from wider-area, full-filter medium band data from CANUCS, which is already providing evidence of more rapid evolution in the UV luminosity function compared to broad band surveys \citep[][]{willott24}.



\section{Science Demonstration: clumpy, offset line emission in reionization-era galaxies}
\label{sec:demo}

In this section, we demonstrate how MegaScience imaging can be used to create maps of both line and continuum emission in galaxies, focusing on H$\beta$+[OIII] line emission in three spectroscopically-confirmed galaxies at $z\sim6.3$. Spatially-extended ionized regions have begun to be observed at $z\gtrsim6$ even in broad-band NIRCam imaging, pointing towards the the importance of ionized galaxy outflows at high redshfit \citep[e.g.,][]{fujimoto22,zhang23}. By more narrowly sampling the line-emitting region, medium band imaging is a powerful tool to accelerate these studies of emission-line morphologies in the era of  reionization and cosmic dawn. 
Unlike the color images shown in Section~\ref{sec:resolved_science}, the images we show here are PSF-matched to the F444W filter using the same methodology described in \citet{weaver24}.

\begin{figure*}
    \centering
    \includegraphics[width=.95\textwidth]{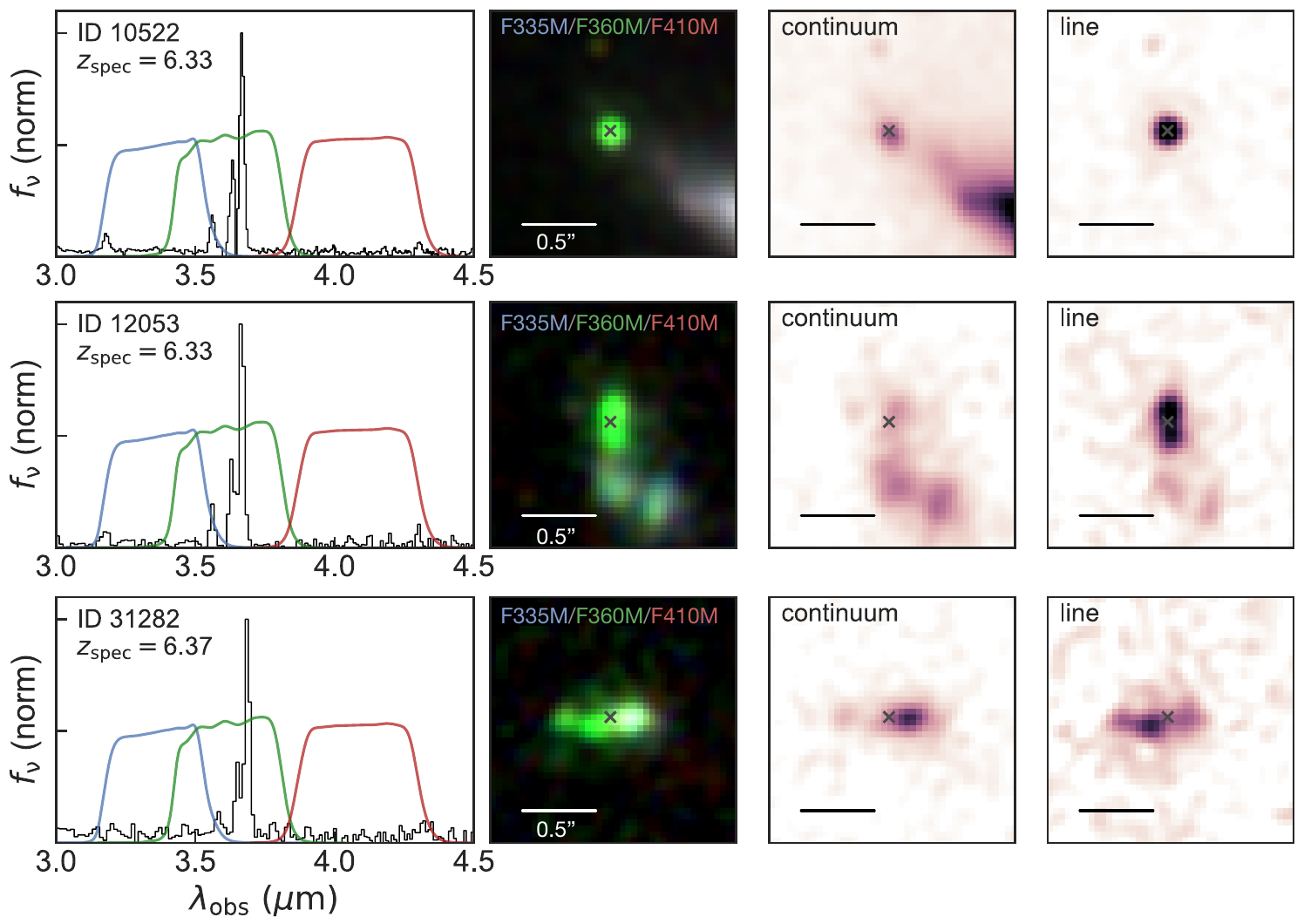}
    \caption{Left: UNCOVER spectroscopy of three $z\sim6.3$ objects with bright [OIII] emission; blue and red curves indicate the medium-band filters used to represent the continuum, while the green curve shows the filter containing the strong line. Center: three-color PSF-convolved images of each source. Right: continuum and line maps for each source. While ID 10522 appears point-like in both continuum and line emission, both 12053 and 31282 exhibit complex morphologies, with three separate clumps at the same redshift, and clear spatial offsets between continuum and line emission.}
    \label{fig:science_demo}
\end{figure*}

We select three example galaxies from a preliminary UNCOVER spectroscopic redshift catalog (S. Price et al., in prep) which have clear H$\beta$+[OIII] emission at $\sim3.6\mu$m, in the center of the F360M filter. We take the neighboring medium-band filters -- F335M on the blue side, and F410M on the red side -- to represent the continuum emission in these galaxies. The left column of Figure~\ref{fig:science_demo} shows the UNCOVER spectra (black) as well as the filter curves for the F335M (blue), F360M (green), and F410M (red) filters. The spectrum confirms that the continuum is relatively flat across the 3-4$\mu$m range, and the continuum filters are free of any major emission lines. 
The central column shows color images of each of the objects using these three medium-band filters; the top source resembles a point source, while the other two have complex and clumpy morphologies.

We then use these three-color images to make simple maps of the continuum and line emission in these galaxies. We take the average of the blue and red continuum bands (F335M and F410M) as the continuum map, shown in the central column. The line map, shown in the rightmost column, is constructed by subtracting the continuum map from the F360M image, which contains both line and continuum emission from the galaxy. This method of deriving the spatial distribution of the continuum may be too simple for a randomly-selected galaxy in our sample --- it could easily be biased by other lines contaminating the ``continuum" bands or a strong spectral slope. However, because the spectra in Fig~\ref{fig:science_demo} show that the continuum is relatively flat and line-free across these filters, this simple method of spatially distinguishing line and continuum emission is likely sufficient for these three sources. 

A grey ``x" marks the center of the color images and continuum and line maps. While the top source, ID 10522, appears to be compact in all maps (with a size consistent with being an unresolved source), IDs 12053 and 31282 have clear differences in the continuum and line maps. In both cases, the color image shows three distinct clumps. In ID 12053, the central clump is significantly brighter in line emission while the two southern clumps are brighter in continuum; in ID 21382, the central clump is brightest in line emission while the western clump is brightest in continuum emission. We note that ID 12053 was detected in [CII] 158$\mu$m by the DUALz survey \citep{fujimoto23}, with a $\sim300$km/s offset between the [CII] 158$\mu$m and [OIII] 88$\mu$m emission. \citet{fujimoto23} discusses that this kinematic offset may be due to complex gas dynamics due to an ongoing merger, an interpretation that is further supported by our observation of different morphologies between the [OIII] line emission and $\sim3\mu$m continuum emission in this object.

The high sensitivity and spatial resolution ($\sim0\farcs04-0\farcs15$) achieved by the NIRCam 1-5$\mu$m filters in this study have revealed the detailed structures of the stellar component and ionized gas in the rest-optical wavelengths, unveiling the complex and clumpy morphologies of galaxies at $z\sim6$. A growing number of studies have identified similarly resolved sources and/or sources with very nearby companions at $z\gtrsim6$ \citep[e.g.,][]{matthee23_eiger,hainline23,yan23,kartaltepe23,witten24}. These findings are consistent with pre-JWST observations finding a trend of increasing clumpiness \citep[e.g.,][]{livermore12,livermore15,shibuya15} and merging fraction \citep[e.g.,][]{romano21} towards higher redshifts. Recent JWST lensing observations have also revealed high clumpiness in galaxies at $z\gtrsim6$ \citep[e.g.,][]{vanzella23,adamo24,mowla24,fujimoto24}. While mergers play a crucial role in mass assembly, the high clumpiness within individual galaxies suggests that active star formation may be driven at least in part by disk instability \citep[e.g.,][]{tadaki18,fujimoto24}. Our observations demonstrate the complex assembly histories of galaxies even at very early cosmic times.

\section{Discussion \& Conclusions}

In this paper, we describe the ``Medium Bands, Mega Science" survey, a survey of Abell 2744 using {\it all} medium- and broad-band JWST/NIRCam filters, and publicly release reduced images and a photometric catalog. With 20 bands of JWST photometry spanning the full 0.7-5$\mu$m range, our data are able to dramatically improve photometric redshift estimates and stellar population fitting. As shown in Figure~\ref{fig:seds}, the MegaScience data is able to precisely constrain break strengths and stellar ages, mitigate confusion between the Balmer and Lyman breaks, and disentangle the contribution of line and continuum emission to broad-band photometry. On a survey-wide basis, comparison to high-quality prism redshifts from UNCOVER (S. Price et al. in prep.) shows that MegaScience photometric redshifts have a factor of $\sim3$ lower scatter and a factor of $\sim2$ fewer catastrophic outliers than the UNCOVER broad-band photometric redshifts. These factors of 2-3 improvement in photometric redshifts will propagate down to improvements in other stellar population properties such as stellar masses, star formation rates, and dust contents; a full exploration of the stellar population properties of galaxies in the survey will be presented in B. Wang et al. in prep.

In addition to improving our understanding of the spatially-integrated properties of galaxies in our survey, the high spatial resolution of this data combined with lensing magnifications from the cluster itself enable us to spatially-resolve galaxy growth, mapping up to $\sim500$pc physical scales across more than ten gigayears of cosmic history. These observations add to an ongoing treasury of deep data across this field: Abell 2744 has now been observed using nearly every instrument and mode on JWST, except MIRI. These data will allow us to compare and calibrate NIRCam imaging, NIRCam grism, and NIRSpec prism and medium-resolution spectroscopy, and understand the formation and evolution of the earliest galaxies.

We publicly provide full reduced images as well as our photometric catalog and both \texttt{eazy-py} and \texttt{Prospector} photometric redshift catalogs on our website, \href{https:/jwst-uncover.github.io/megascience/}{jwst-uncover.github.io/megascience} as well as Zenodo, \href{https://zenodo.org/doi/10.5281/zenodo.8199802}{zenodo.org/doi/10.5281/zenodo.8199802}. Future image and photometric catalog releases will be updated on our website. Our team will also publicly provide \texttt{Prospector} catalogs of stellar population properties (B. Wang et al. in prep.). 



\label{sec:discussion}

\acknowledgements 
KAS thanks Shelly Meyett, the MegaScience Program Coordinator, for invaluable assistance designing WOPR 88967 and ensuring that this program was fully observed. This proposal was conceived of and developed at the International Space Science Institute (ISSI) in Bern, through ISSI International Team project \#562.  
This work is based on observations made with the NASA/ESA/CSA James Webb Space Telescope. The raw data were obtained from the Mikulski Archive for Space Telescopes at the Space Telescope Science Institute, which is operated by the Association of Universities for Research in Astronomy, Inc., under NASA contract NAS 5-03127 for \textit{JWST}. These observations are associated with JWST Cycle 2 GO program \#4111, and this project has gratefully made use of a large number of public JWST programs in the Abell 2744 field including JWST-GO-2641, JWST-ERS-1324, JWST-DD-2756, JWST-GO-2883, JWST-GO-3538, and JWST-GO-3516. Support for program JWST-GO-4111 was provided by NASA through a grant from the Space Telescope Science Institute, which is operated by the Associations of Universities for Research in Astronomy, Incorporated, under NASA contract NAS5-26555. 

The Cosmic Dawn Center is funded by the Danish National Research Foundation
(DNRF) under grant \#140. PD acknowledges support from the NWO grant 016.VIDI.189.162 (``ODIN") and warmly thanks the European Commission's and University of Groningen's CO-FUND Rosalind Franklin program.
This work has received funding from the Swiss State Secretariat for Education, Research and Innovation (SERI) under contract number MB22.00072, as well as from the Swiss National Science Foundation (SNSF) through project grant 200020\_207349.
The work of CCW is supported by NOIRLab, which is managed by the Association of Universities for Research in Astronomy (AURA) under a cooperative agreement with the National Science Foundation.
TBM was supported by a CIERA Fellowship.
The BGU lensing group acknowledges support by Grant No. 2020750 from the United States-Israel Binational Science Foundation (BSF) and Grant No. 2109066 from the United States National Science Foundation (NSF); by the Ministry of Science \& Technology, Israel; and by the Israel Science Foundation Grant No. 864/23.
YF acknowledges support from JSPS KAKENHI Grant Number JP22K21349 and JP23K13149. RP and DM acknowledge funding from JWST-GO-02561.013 and JWST-GO-04111.035.

\software{astropy \citep{astropy2013, astropy2018}, grizli \citep{brammer14}, Seaborn \citep{waskom17}, WebbPSF \citep{perrin14}}

\bibliographystyle{aasjournal}
\bibliography{all}

\end{document}